\documentclass[showpacs,aps,prl,reprint,superscriptaddress,preprintnumbers,nolinenumbers]{revtex4-1}

\usepackage{amsmath}
\usepackage{amssymb}
\usepackage{graphicx}
\usepackage{dcolumn}
\usepackage{color}
\usepackage{xcolor}
\usepackage{endnotes}
\usepackage{bm}
\usepackage{epsfig}
\usepackage{array}

\begin{document}



\title{{Detection of hole pockets in the candidate type-II Weyl semimetal MoTe$_2$ from Shubnikov-de Haas quantum oscillations}}

\author{Y. J. Hu}
\affiliation{Department of Physics, The Chinese University of Hong Kong, Shatin, Hong Kong}
\author{W. C. Yu}
\email[]{wingcyu@cityu.edu.hk}
\affiliation{Department of Physics, The Chinese University of Hong Kong, Shatin, Hong Kong}
\affiliation{Department of Physics, City University of Hong Kong, Kowloon, Hong Kong}
\author{Kwing To Lai}
\affiliation{Department of Physics, The Chinese University of Hong Kong, Shatin, Hong Kong}
\author{D. Sun}
\author{F. F. Balakirev}
\affiliation{National High Magnetic Field Laboratory, Los Alamos National Laboratory, Los Alamos, New Mexico 87545, USA}
\author{W.~Zhang}
\author{J.~Y.~Xie}
\author{K. Y. Yip}
\author{E.~I.~Paredes~Aulestia}
\affiliation{Department of Physics, The Chinese University of Hong Kong, Shatin, Hong Kong}
\author{Rajveer~Jha}
\author{Ryuji~Higashinaka}
\author{Tatsuma D. Matsuda}
\affiliation{Department of Physics, Tokyo Metropolitan University, Hachioji, Tokyo 192-0397, Japan}
\author{Y.~Yanase}
\affiliation{Department of Physics, Kyoto University, Kyoto 606-8502, Japan}
\author{Yuji Aoki}
\affiliation{Department of Physics, Tokyo Metropolitan University, Hachioji, Tokyo 192-0397, Japan}
\author{Swee K. Goh}
\email[]{skgoh@cuhk.edu.hk}
\affiliation{Department of Physics, The Chinese University of Hong Kong, Shatin, Hong Kong}

\date{24 January 2020}

\begin{abstract}
The bulk electronic structure of $T_d$-MoTe$_2$ features large hole Fermi pockets at the Brillouin zone center ($\Gamma$) and two electron Fermi surfaces along the $\Gamma-X$ direction. However, the large hole pockets, whose existence has important implications for the Weyl physics of $T_d$-MoTe$_2$, has never been conclusively detected in quantum oscillations. This raises doubt about the realizability of Majorana states in $T_d$-MoTe$_2$, because these exotic states rely on the existence of Weyl points, which originated from the same band structure predicted by density functional theory (DFT).
Here, we report an unambiguous detection of these elusive hole pockets via Shubnikov-de Haas (SdH) quantum oscillations. At ambient pressure, the quantum oscillation frequencies for these pockets are 988~T and 1513~T, when the magnetic field is applied along the $c$-axis. The quasiparticle effective masses $m^*$ associated with these frequencies are 1.50~$m_e$ and 2.77~$m_e$, respectively, indicating the importance of Coulomb interactions in this system. 
We further measure the SdH oscillations under pressure. At 13~kbar, we detected a peak at 1798~T with $m^*$ = 2.86~$m_e$. Relative to the oscillation data at a lower pressure, the amplitude of this peak experienced an enhancement, which can be attributed to the reduced curvature of the hole pockets under pressure. Combining our experimental data with DFT + $U$ calculations, where $U$ is the Hubbard parameter, our results shed light on why these important hole pockets have not been detected until now.
\end{abstract}


\maketitle


$T_d$-MoTe$_2$, like its prominent relative WTe$_2$, is an important member of the transition metal dichalcogenide family exhibiting interesting electronic properties \cite{Yan2017}. $T_d$-MoTe$_2$ has been predicted to be a type-II Weyl semimetal, in which the Weyl points arise from linear touching points at the boundary between electron and hole pockets \cite{Sun2015, Soluyanov2015, Wu2016prb, Deng2016, Jiang2017nc, Wang2016, Aryal2019}. Furthermore, $T_d$-MoTe$_2$ exhibits a non-saturating, quadratic extremely large magnetoresistance (XMR) at low temperatures \cite{Keum2015}, which can at least be partially attributed to a perfect compensation between electron and hole carriers \cite{Zhou2016prb, Lee2018, Chen2016a}. 
Additionally, $T_d$-MoTe$_2$ undergoes a superconducting transition at $T_c\sim 0.1$~K and a structural transition to the $1T'$ phase above $T_s\sim$ 263~K at ambient pressure \cite{Qi2016}.  Under pressure, $T_s$ is rapidly suppressed, extrapolating to 0 K at $\sim$11~kbar \cite{Takahashi2017, Heikes2018, Lee2018, Hu2019}, while $T_c$ is rapidly enhanced. This makes MoTe$_2$ interesting because of the ability to access both the $T_d$ and $1T'$ phase with a moderate pressure. At 15~kbar ($1T'$ phase), the 2D superconductivity has been reported \cite{Hu2019}.

For a thorough discussion of the intriguing properties of MoTe$_2$, it is imperative to understand its electronic structure. Several early angle-resolved photoemission spectroscopy (ARPES) measurements have claimed to observe Weyl points and Fermi arcs in $T_d$-MoTe$_2$, and the measured bulk Fermi surfaces are in broad agreement with the prediction of density functional theory (DFT) \cite{Deng2016, Jiang2017nc, Crepaldi2017, Tamai2016, Weber2018}. However, an agreement between quantum oscillations and DFT has been significantly less satisfying: the large hole pockets 
predicted in DFT \cite{Rhodes2017} did not show up in quantum oscillation data \cite{Rhodes2017, Qi2016, Zhou2016prb, Chen2018prb, Luo2016, Zhong2018, Liu2019}. To match the experimental data with the calculation, an {\it ad hoc} shift of the valance bands relative to the Fermi energy was performed \cite{Rhodes2017}. Unfortunately, such a band shifting eliminates the type-II Weyl points, raising doubt regarding the true topological nature of $T_d$-MoTe$_2$. Although Majorana states due to the interplay between topological semimetal and superconductivity have been proposed \cite{AChen2016, AChen2017}, their realizability in $T_d$-MoTe$_2$ becomes doubtful with the elimination of Weyl points. On the other hand, the absence of hole pockets would present significant challenges to the understanding of $T_d$-MoTe$_2$. For instance, it would be impossible to reconcile the absence of hole pockets with the interpretation of the XMR based on electron-hole carrier compensation scenario.

Recently, it has been pointed out that if Coulomb interaction is included in DFT calculations, a better agreement with ARPES data can be achieved \cite{Aryal2019, Xu2018}. Furthermore, in these DFT + $U$ calculations, where $U$ is the on-site Coulomb interaction of Mo $4d$ electrons, the shift of energy bands similar to the operation attempted in Ref.~\cite{Rhodes2017} can be reproduced without the elimination of the type-II Weyl points. With an increasing $U$, the hole pockets are shrinking. For instance, the frequency of the biggest hole pocket when the magnetic field ($B$) is applied along the $c$-axis decreases monotonically from $\sim$2000~T at $U=0$ to $\sim$1500~T at $U=4$~eV \cite{Aryal2019}. 
Hence, even with the inclusion of a sizeable $U$, a distinct quantum oscillation frequency larger than $\sim$1000~T should still appear, which had not been detected experimentally \cite{Rhodes2017, Qi2016, Zhou2016prb, Chen2018prb, Luo2016, Zhong2018, Liu2019}.  
Therefore, the observation of this particular frequency when $B\parallel c$ has been long awaited and it can be regarded as the ``smoking gun" evidence for the correctness of DFT predictions, and it serves to pinpoint a value of $U$. This observation would also represent an unambiguous detection of the hole pockets, which is clearly crucial for understanding MoTe$_2$ physics. In this manuscript, we present our Shubnikov-de Haas (SdH) oscillations from high quality MoTe$_2$ single crystals. With $B\parallel c$,
we successfully detect new frequencies that can be assigned to the hole pockets, in excellent agreement with DFT + $U$ calculations. Moreover, our data show that with an increasing pressure, the hole pockets become more two-dimensional and cylindrical, explaining why the SdH oscillations are hard to detect at ambient pressure.

\begin{figure}[!t]\centering
      \resizebox{9cm}{!}{
              \includegraphics{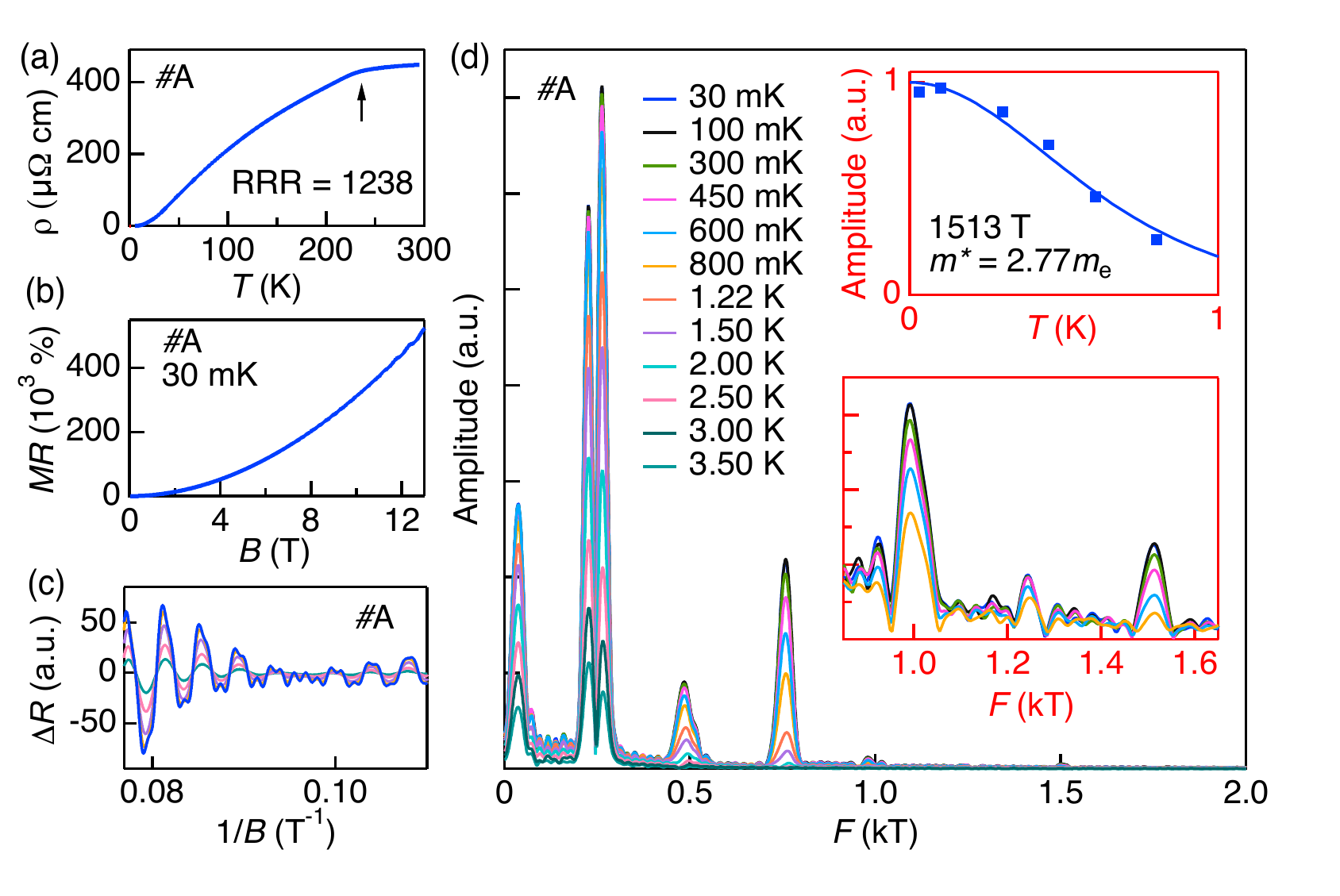}}                				
              \caption{\label{fig1} Electrical transport data collected on sample \#A at ambient pressure. (a) The temperature dependence of the resistivity upon cooling. The arrow indicates the anomaly associated with the structural transition. (b) The MR at 30~mK with $B\parallel c$. (c) SdH oscillations at several representative temperatures. (d) Quantum oscillation spectra for $B\parallel c$ at temperatures ranging from 30 mK to 3.5 K. The FFT was performed from 8~T to 13~T. The lower inset shows the FFT spectra from 0.85~kT to 1.65~kT at 30 mK, 100~mK, 300~mK, 450~mK, 600~mK and 800~mK. To show the high-frequency peaks clearly, the FFT was performed from 9.5~T to 13~T. The upper inset plots the amplitudes of the 1513~T peak against the temperature.}
\end{figure}

Single crystals of MoTe$_2$ were synthesized by the Te-flux method as described in Supplemental Material \cite{SUPP}.
Three thin MoTe$_2$ samples with a thickness of around 10 $\mu$m were cleaved from the same crystal:  samples $\#$A and $\#$C for ambient pressure, and sample $\#$B for high pressure measurements. $\#$C was measured at The National High Magnetic Field Laboratory (NHMFL) in Tallahassee. These thicknesses guarantee that our samples are in the bulk limit. The electrical resistivity up to 14~T was measured by a four-terminal configuration either in a Physical Property Measurement System by Quantum Design (down to 2~K) or a dilution fridge by Bluefors (down to 30~mK). Hydrostatic pressure was provided by a piston-cylinder clamp cell with glycerin as the pressure transmitting medium. The pressure value was determined resistively by the superconducting transition of Pb. DFT + $U$ calculations were performed, with details provided in Supplemental Material \cite{SUPP}.


We begin by a presentation of our ambient pressure data. Figure~\ref{fig1}(a) shows the zero-field $\rho(T)$ of sample \#A at ambient pressure, collected on cooling. An anomaly in $\rho(T)$ is observed at 236~K at ambient pressure, corresponding to the first-order structural transition. The residual resistivity ratio [RRR=$\rho$(300K)/$\rho$(2K)] is 1238, indicating superior sample quality. Figure~\ref{fig1}(b) shows the magnetoresistance (MR) at 30~mK with $B\parallel c$, reaching a value of $\sim523,100$\% at 13~T, manifesting again the high quality of our sample. With the large magnetoresistance background removed, clear Shubnikov-de Haas oscillations can be seen (Fig.~\ref{fig1}(c)). In Fig.~\ref{fig1}(d), we display the SdH spectra obtained from Fast Fourier transform (FFT) for a wide range of temperatures. Our SdH spectra for $B\parallel c$ are considerably richer than most existing data in the literature thus far \cite{Rhodes2017, Qi2016, Zhou2016prb, Chen2018prb, Luo2016, Zhong2018, Liu2019}: $F_\alpha=226$~T and $F_\beta=263$~T are the same as previously reported peaks \cite{Rhodes2017, Qi2016, Zhou2016prb, Chen2018prb, Luo2016, Zhong2018, Liu2019}, and they correspond to the electron pockets. However, we also detect more frequencies, including $F_\gamma=484$~T, $F_\delta=758$~T, $F_{2\beta}=518$~T.

A long-standing mystery in the studies of MoTe$_2$ electronic structure concerns the absence of any quantum oscillation frequencies above 1000~T with $B\parallel c$ in experiments. When we examine our SdH data closely, we discover new frequencies at 988~T and 1513~T. As shown in the lower inset of Fig.~\ref{fig1}(d), the peaks are weak but well above the noise floor. The temperature dependence of the amplitudes follows the prediction of Lifshitz-Kosevich theory, enabling the extraction of sheet-resolved effective masses $m^*$. For the peak at 1513~T, $m^*=(2.77\pm0.15)m_e$ (see the upper inset of Fig.~\ref{fig1}(d)). For the peak at 988~T, $m^*=(1.50\pm0.03)m_e$  \cite{SUPP}. Note that there is a much weaker peak at 1244~T, which could be an artifact of the experiment. The frequency of the 1513~T peak is nearly $2F_\delta$. However, the effective mass of $\delta$ is $(1.99\pm0.06)m_e$ \cite{SUPP}, which is significantly more than half of $2.77m_e$. Hence, we can exclude the possibility that the 1513~T peak is the second harmonic of $\delta$. 
We further varied the field angle at 30~mK to track the angular dependence of the new found SdH frequencies (see Supplemental Material \cite{SUPP}). The comparison between our data and the DFT+$U$ calculations enables us to definitively associate the frequencies at 988~T and 1513~T with the large hole pockets in $T_d$-MoTe$_2$. Besides, the rather large $m^*$ highlights the importance of electron-electron correlations, and justifies the inclusion of Hubbard $U$ in the calculations. 

\begin{figure}[!t]\centering
       \resizebox{8.5cm}{!}{
              \includegraphics{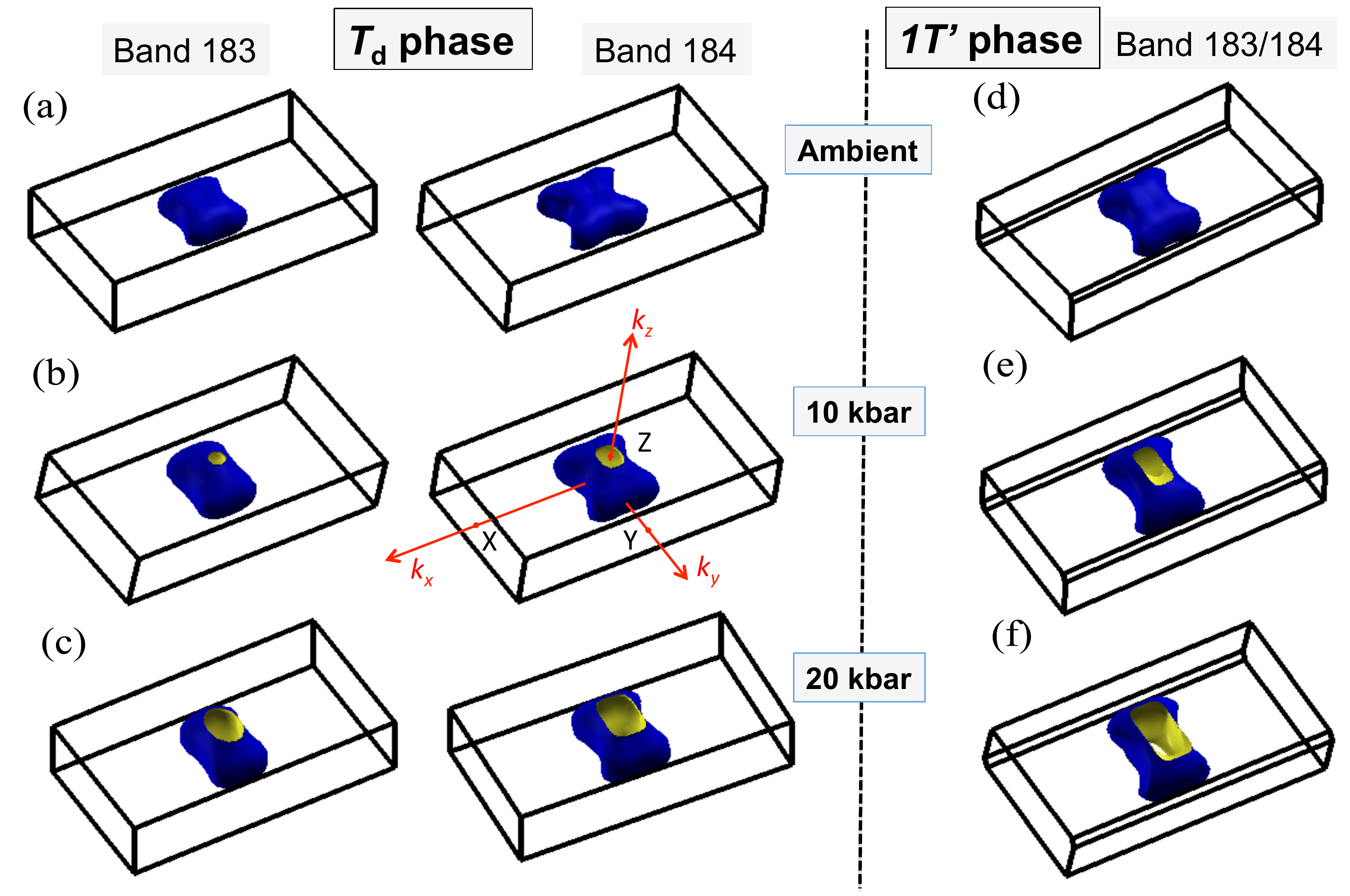}}                				
              \caption{\label{fig2} Hole Fermi surface(s) centered at the $\Gamma$ point of the Brillouin zone at ambient pressure, 10~kbar and 20~kbar, calculated with $U=3$~eV. (a)-(c) displays the results in the $T_d$ phase. In the $1T'$ phase ((d)-(f)), Band 183 and Band 184 become degenerate.}
\end{figure}

Figure~\ref{fig2}(a) displays the calculated hole Fermi surfaces that are of central interest in this Letter. At ambient pressure, in the $T_d$ phase, they are large and closed, three-dimensional pockets.
Since our angular dependence of SdH frequencies indeed exhibits a good agreement with the calculations with $U=3$~eV \cite{SUPP}, we fix $U$ to 3~eV. We point out that the curvature of the hole pockets along the $k_z$ direction is large. We speculate that this relatively large curvature, together with the large $m^*$, are responsible for the difficulty in detecting these frequencies. 

The curvature of the hole pockets can be varied by applying hydrostatic pressure. To see how the curvature evolves, we perform calculations at 10~kbar and 20~kbar, as shown in Figs.~\ref{fig2}(b),(c) and (e),(f). Although MoTe$_2$ is in the $1T'$ phase at 20~kbar \cite{Hu2019}, calculations using the space group of both the $T_d$ and $1T'$ phases were performed to theoretically investigate the curvature change with an increasing pressure. 
In the $1T'$ phase, Bands 183 and 184 are degenerate because of the restoration of the inversion symmetry. For the calculations in both $1T'$ and $T_d$ phases, the hole pocket(s) expand in size with an increasing pressure, and they progressively touch the Brillouin zone boundary along the $k_z$ direction. Hence, we conclude that the Fermi surfaces become more cylindrical and the curvature along the $k_z$ direction {\it decreases} under pressure.

The effect of the Fermi surface curvature on quantum oscillation amplitude is usually incorporated by slicing a Fermi surface into thin slabs that are perpendicular to the magnetic field \cite{Shoenbergbook}. Without loss of generality, we take ${\bf B}=B\hat{z}$. The thickness and the cross-sectional area of the slabs become $d\kappa_z$ and $S(\kappa_z)$, respectively, where the $\kappa_z$ axis is parallel to the $k_z$ axis. To get the total signal contributed by these slices, we need to integrate along $\kappa_z$:
\begin{equation}
\int d\kappa_z \sum_{p=1}^{\infty} \cos\left[2\pi p \left(\frac{\hbar S(\kappa_z)}{2\pi e B}-\gamma\right)\right].
\end{equation}
where $p$ is the harmonic index and $\gamma$ is a phase factor. Taking the extremal area $S_{\rm ext}$ to be at $\kappa_z=0$, and focusing only on the range of $\kappa_z$ in the vicinity of the extremal area, the integral can be approximated as \cite{Shoenbergbook}
\begin{equation}
\left(\left|\frac{\partial^2 S}{\partial \kappa_z^2}\right|_{\kappa_z=0}\right)^{-\frac{1}{2}}\sum_{p=1}^{\infty}{p^{-\frac{1}{2}}}\cos\left[2\pi p \left(\frac{F}{B}-\gamma\right)\pm\frac{\pi}{4}\right].
\end{equation}
where $F=\hbar S_{\rm ext}/(2\pi e)$ is the quantum oscillation frequency corresponding to the extremal area. The second derivative, which characterizes the curvature of the Fermi surface at $\kappa_z=0$, enters as the denominator of the prefactor. Therefore, a larger curvature will result in a smaller quantum oscillation amplitude.
\begin{figure}[!t]\centering
       \resizebox{8.5cm}{!}{
              \includegraphics{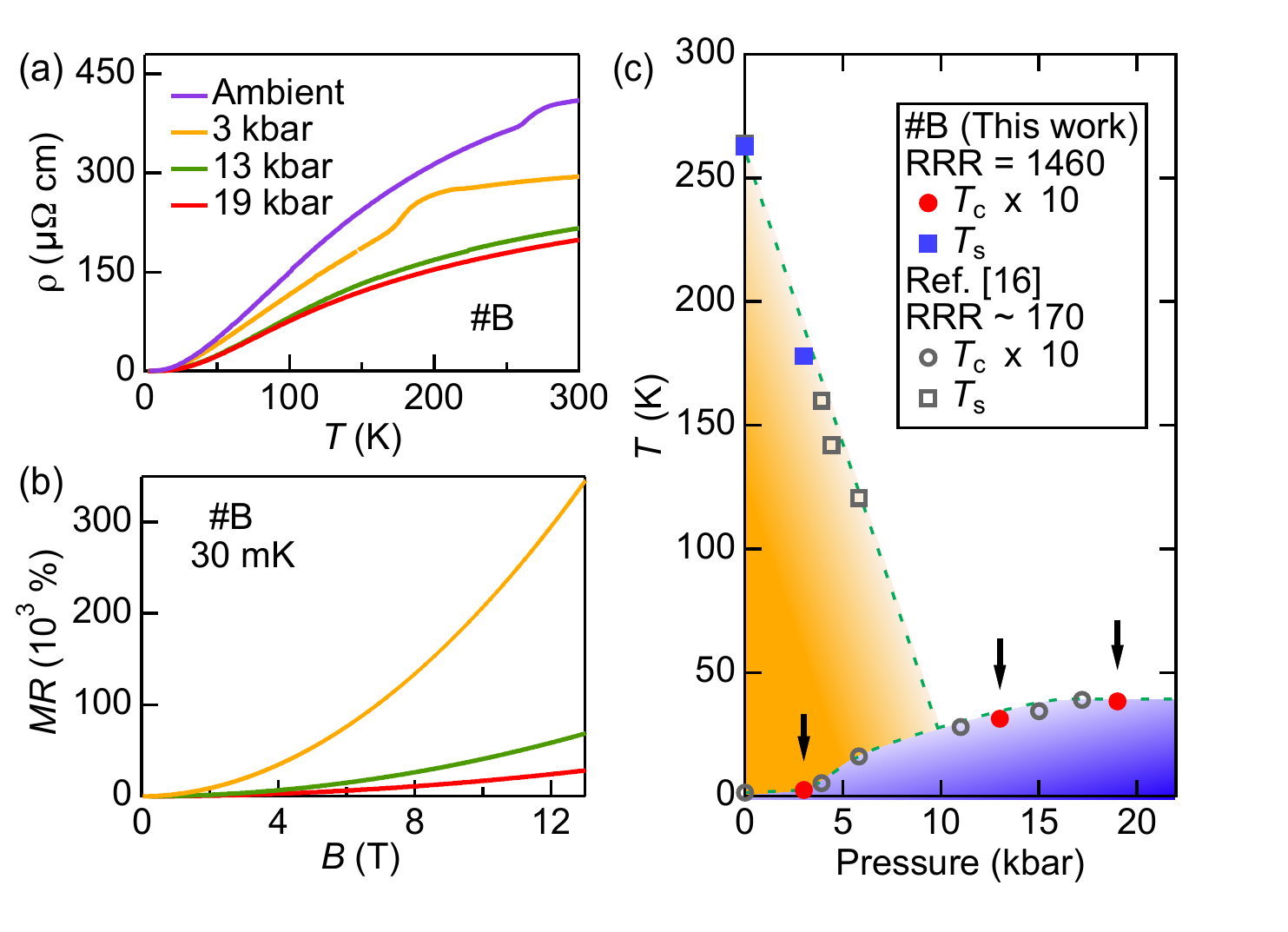}}                				
              \caption{\label{fig3} (a) Temperature dependence of the resistivity at different pressures for sample \#B, collected on warming up. Sample \#B has a RRR of 1460 at ambient pressure. (b) The field dependence of the MR with $B\parallel c$ at 30~mK at 3~kbar, 13~kbar and 19~kbar. (c) Temperature-pressure phase diagram. $T_s$ is the structural transition temperature on warming, $T_c$ is the superconducting transition temperature. Closed symbols are from the current work, open symbols are from Ref.~\cite{Hu2019}. The arrows indicate the pressure points for the SdH studies of sample \#B.}
\end{figure}

\begin{figure}[!t]\centering
   \resizebox{8.5cm}{!}{
           \includegraphics{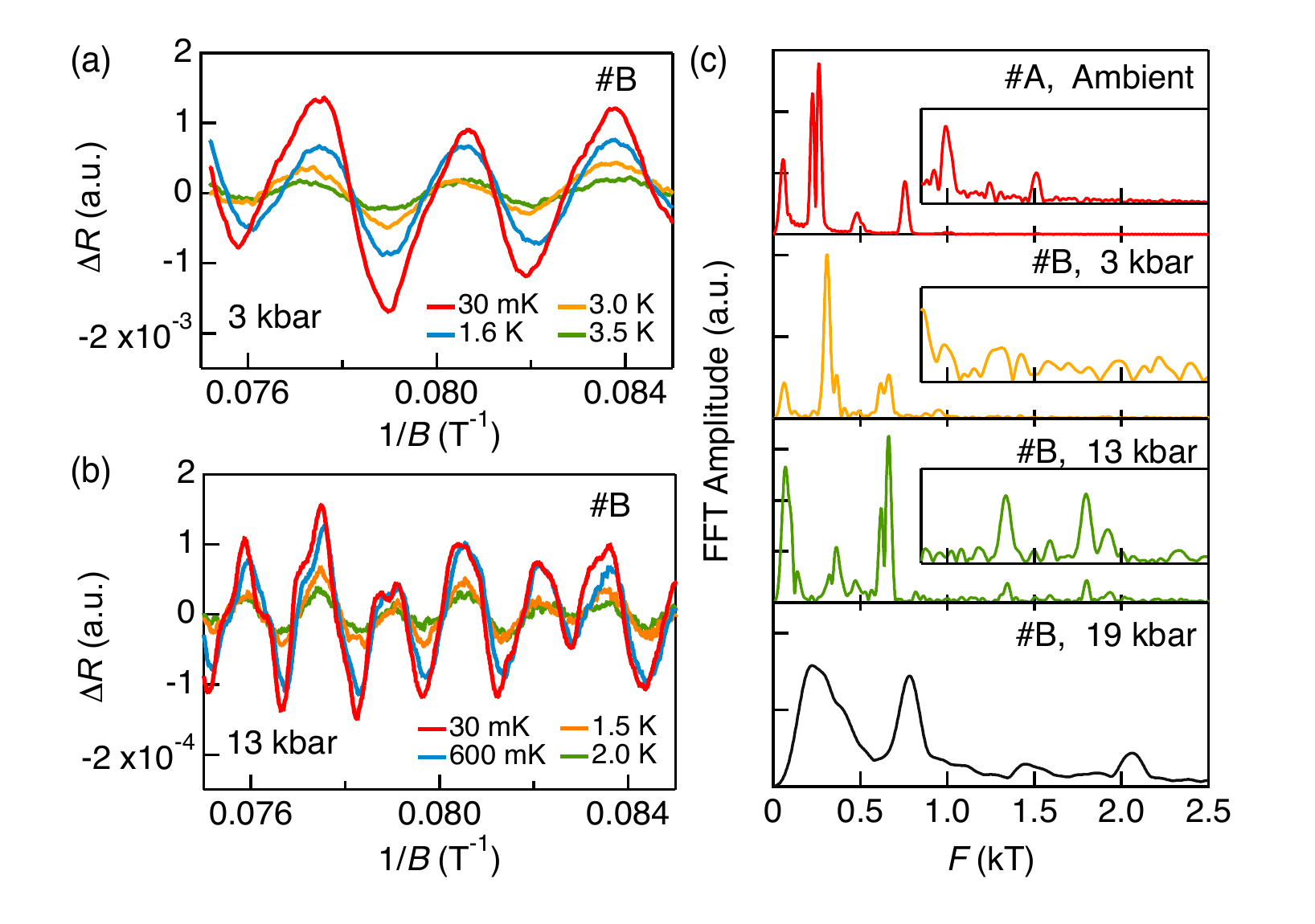}}                				
          \caption{\label{fig4} SdH oscillations for MoTe$_2$ at (a) 3~kbar and (b) 13~kbar. (c) The pressure evolution of SdH spectra at 30~mK. For the spectra shown in the main panels, the FFT was performed using data between 8~T and 13~T for ambient pressure, 3~kbar and 13~kbar, and between 11.3~T and 13.3~T for 19 kbar. The magnetic field is applied along the $c$-axis. The insets exhibit the SdH spectra from 850~T to 2500~T, obtained with FFT performed between 9.5~T and 13~T. }
\end{figure}

Since our DFT + $U$ calculations have established that the curvature of the hole pockets along the $k_z$ direction decreases under pressure, the SdH oscillation amplitudes should be enhanced. To test this hypothesis, we conduct a series of high-pressure electrical transport experiments.
Figures~\ref{fig3}(a) and (b) exhibit $\rho(T)$ and the field dependence of MR at 30~mK, respectively. The rapid suppression of MR under pressure is qualitatively similar to the recent study using crystals with a lower RRR of $\sim$170 \cite{Hu2019}.
However, the MR(13~T, 30~mK) at 19~kbar for the present work remains large, having a value of 28,300\%, much larger than the value of 1,000\% at 17~kbar in Ref. \cite{Hu2019}. From our electrical transport data, we can extract $T_s$ and $T_c$, which are shown in Fig.~\ref{fig4}(c). The pressure dependence of $T_s$ and $T_c$ from samples with RRR of $\sim$170 are also included, which agree nicely with the present work. The overall smooth variation of $T_s(p)$ and $T_c(p)$ rule out any strong dependence on RRR.

The large MR offers the prospect of detecting quantum oscillations under pressure. Clear SdH oscillations can be seen at 3~kbar and 13~kbar, as displayed in Figs.~\ref{fig4}(a) and (b).
The pressure-dependent SdH spectra at 30 mK with $B\parallel c$ are shown in Fig.~\ref{fig4}(c). The ambient pressure SdH spectrum is the same as the one shown in Fig.~\ref{fig1}(d). We stress that the pressure work was conducted on sample $\#$B, which is from the same batch as $\#$A used for ambient pressure work. At 3~kbar, several peaks below 1000~T can be resolved, and an intense peak at 309~T is most likely related to $F_\alpha=226$~T observed at ambient pressure. We do not observe any peak higher than 1000~T at 3~kbar, as can be seen from the inset. This can be understood because sample \#B has a lower signal-to-noise ratio than \#A. In this low pressure range, the high-frequency oscillations are still weak and they can easily be buried in the noise.

Upon increasing pressure to 13~kbar, an interesting SdH spectrum is obtained. While the strongest peak continues to shift toward the higher frequency end, two new peaks appear: one at 1342~T and another at 1798~T. The frequency of the 1342~T peak is nearly double that of the most intense peak (661~T). 
Furthermore, its effective mass is $(3.11\pm 0.17)m_e$, which is exactly twice the effective mass associated with the 661~T peak, $(1.55\pm 0.03)m_e$ \cite{SUPP}.
Hence, the 1342~T peak is the second harmonic. However, the 1798~T peak is not the integer multiple of any peak at a lower frequency. Hence, we conclude that this peak is indeed independent. Finally, we note that the effective masses at 13~kbar are significantly larger than the ambient pressure values \cite{SUPP}: $m^*$ is $(2.86\pm 0.19)m_e$,  $(1.61\pm 0.11)m_e$ and $(1.46\pm 0.03)m_e$ for the 1798~T, 324~T and 621~T peaks, respectively.  The enhanced effective masses further suggest the importance of Coulomb interactions under pressure.

The observation of a SdH peak with a larger amplitude with an increasing pressure is extraordinary. Empirically, SdH amplitudes usually decrease under pressure \cite{Cai2015, Zhang2017}. For instance, the strongest peak at 13~kbar (661~T) is 12.5 times weaker than that at 3~kbar (309~T). However, the unusual growth of the 1798~T peak at 13~kbar can be nicely attributed to the decrease of the curvature of the hole Fermi surfaces discussed above. Furthermore, since MoTe$_2$ is already in the $1T'$ phase at 13~kbar \cite{Hu2019}, Bands 183 and 184 become degenerate due to the recovered inversion symmetry. The multiplicity is expected to further enhance the quantum oscillation signals. 
Therefore, combining the calculations and the experimental results, we conclude that the 1798~T peak at 13~kbar is related to the 988~T and 1513~T peaks at ambient pressure, which further strengthens our conclusion of detecting hole pockets. 

At 19~kbar, the SdH oscillations fade away. Similar weakening of the SdH amplitudes was also reported in WTe$_2$ \cite{Cai2015}, which can be intuitively attributed to pressure inhomogeneity and the suppression of MR. Fortunately, with an average over 16 field sweeps under identical experimental conditions, several broad peaks can be resolved. Although the weak SdH signals render it impossible to reliably obtain the effective mass, the spectrum is useful for the discussion of pressure evolution of SdH frequencies. From the spectrum, we identify three peaks located at 780~T, 1450~T, and 2066~T. The first two peaks can be taken as the fundamental and its harmonic. Given the relatively small pressure difference between 19 kbar and 13 kbar, and the similar structure in the SdH spectra, the 2066~T is naturally recognized to be related to the 1798~T peak at 13~kbar. In fact, linearly extrapolating these frequencies to ambient pressure would give a peak at 1217~T, which is in between the two frequencies from the hole Fermi surfaces at ambient pressure. This is expected since the degeneracy is lifted due to the absence of inversion symmetry in the ambient pressure $T_d$ phase.

Taking into account all SdH data collected, the pressure evolution of our data is consistent with the expectation from DFT+$U$ calculations (Fig.~\ref{fig2}). Thus, the pressure studies lend further support to our central claim on the detection of the hole Fermi surfaces in MoTe$_2$. As pointed out by Aryal and Manousakis \cite{Aryal2019}, the inclusion of $U$ does not annihilate all Weyl points in the $T_d$ phase: at least one pair of type-II Weyl points very close to the Fermi level survive. As a corollary, our work might have resolved a lingering issue concerning the relevance of Weyl physics in $T_d$-MoTe$_2$ \cite{Rhodes2017}, and reestablishes $T_d$-MoTe$_2$ as a candidate of type-II Weyl semimetal.

In summary, we have used the Shubnikov-de Haas effect to probe the electronic structure of MoTe$_2$. We successfully detected the large hole pockets with frequencies of 988~T and 1513~T at ambient pressure, which is in the $T_d$ phase. The cyclotron effective masses associated with these frequencies are $(1.50\pm0.03)m_e$ and $(2.77\pm0.15)m_e$, respectively, indicating a non-negligible correlation effect. At 13~kbar, in the $1T'$ phase, we detected one frequency at 1798~T, whose amplitude experienced an enhancement relative to that at 3~kbar. This peak shifts to a higher value at 19 kbar. This trend is consistent with the predictions of the DFT + $U$ calculations, showing that the hole pockets expand under pressure, accompanied by a decreasing curvature along the $k_z$ direction. Thus, at ambient pressure, the large Fermi surface curvature in the vicinity of the extremal orbit and the large effective mass make it challenging to detect the hole pockets. To overcome this challenge, high quality single crystals and a careful optimization of the experimental conditions are indispensable. Finally, as previous DFT + $U$ calculations \cite{Aryal2019} show that at least one pair of type-II Weyl points is preserved, the consistency between our results and the calculations serves to reestablish $T_d$-MoTe$_2$ as a candidate of type-II Weyl semimetal.

\begin{acknowledgments}
We acknowledge technical assistance from Stanley W. Tozer, Scott A. Maier and Bobby Joe Pullum for high field experiments at NHMFL Tallahassee, and fruitful discussions with Qun Niu, Yuk Tai Chan, Ece Uykur, Xi Dai, Kin On Ho, Xiaoyu Guo and Shingo Yonezawa. This work is supported by Research Grants Council of Hong Kong (CUHK 14300418, CUHK 14300117), CUHK Direct Grant (4053345, 4053299), CityU Start-up Grant (No. 9610438), JSPS KAKENHI (JP15H05884, JP18H04225, JP15H05745, JP18H01178, JP18H05227, JP15K05178, JP16F16028 and JP19H01839). A portion of this work was performed at the National High Magnetic Field Laboratory, which is supported by the National Science Foundation Cooperative Agreement No. DMR-1644779 and the State of Florida.

\end{acknowledgments}


\begin{thebibliography}{10}

\bibitem{Yan2017}
B.~Yan and C.~Felser,
\newblock Annu. Rev. Condens. Matter Phys. {\bf 8}, 337 (2017).

\bibitem{Sun2015}
Y.~Sun, S.-C. Wu, M.~N. Ali, C.~Felser, and B.~Yan,
\newblock Phys. Rev. B {\bf 92}, 161107 (2015).

\bibitem{Soluyanov2015}
A.~A. Soluyanov, D.~Gresch, Z.~Wang, Q.~Wu, M.~Troyer, X.~Dai, and B.~A.
  Bernevig,
\newblock Nature {\bf 527}, 495 (2015).

\bibitem{Wu2016prb}
Y.~Wu, D.~Mou, N.~H. Jo, K.~Sun, L.~Huang, S.~L. Bud'ko, P.~C. Canfield, and
  A.~Kaminski,
\newblock Phys. Rev. B {\bf 94}, 121113 (2016).

\bibitem{Deng2016}
K.~Deng, G.~Wan, P.~Deng, K.~Zhang, S.~Ding, E.~Wang, M.~Yan, H.~Huang,
  H.~Zhang, Z.~Xu, J.~Denlinger, A.~Fedorov, H.~Yang, W.~Duan, H.~Yao, Y.~Wu,
  S.~Fan, H.~Zhang, X.~Chen, and S.~Zhou,
\newblock Nat. Phys. {\bf 12}, 1105 (2016).

\bibitem{Jiang2017nc}
J.~Jiang, Z.~K. Liu, Y.~Sun, H.~F. Yang, C.~R. Rajamathi, Y.~P. Qi, L.~X. Yang,
  C.~Chen, H.~Peng, C.-C. Hwang, S.~Z. Sun, S.-K. Mo, I.~Vobornik, J.~Fujii,
  S.~S.~P. Parkin, C.~Felser, B.~H. Yan, and Y.~L. Chen,
\newblock Nat. Commun. {\bf 8}, 13973 (2017).

\bibitem{Wang2016}
Z.~Wang, D.~Gresch, A.~A. Soluyanov, W.~Xie, S.~Kushwaha, X.~Dai, M.~Troyer,
  R.~J. Cava, and B.~A. Bernevig,
\newblock Phys. Rev. Lett. {\bf 117}, 056805 (2016).

\bibitem{Aryal2019}
N.~Aryal and E.~Manousakis,
\newblock Phys. Rev. B {\bf 99}, 035123 (2019).

\bibitem{Keum2015}
D.~H. Keum, S.~Cho, J.~H. Kim, D.-H. Choe, H.-J. Sung, M.~Kan, H.~Kang, J.-Y.
  Hwang, S.~Kim, H.~Yang, K.~J. Chang, and Y.~H. Lee,
\newblock Nat. Phys. {\bf 11}, 482 (2015).

\bibitem{Zhou2016prb}
Q.~Zhou, D.~Rhodes, Q.~R. Zhang, S.~Tang, R.~Sch\"onemann, and L.~Balicas,
\newblock Phys. Rev. B {\bf 94}, 121101 (2016).

\bibitem{Lee2018}
S.~Lee, J.~Jang, S.-I. Kim, S.-G. Jung, J.~Kim, S.~Cho, S.~W. Kim, J.~Y. Rhee,
  K.-S. Park, and T.~Park,
\newblock Sci. Rep. {\bf 8}, 13937 (2018).

\bibitem{Chen2016a}
F.~C. Chen, H.~Y. Lv, X.~Luo, W.~J. Lu, Q.~L. Pei, G.~T. Lin, Y.~Y. Han, X.~B.
  Zhu, W.~H. Song, and Y.~P. Sun,
\newblock Phys. Rev. B {\bf 94}, 235154 (2016).

\bibitem{Qi2016}
Y.~Qi, P.~G. Naumov, M.~N. Ali, C.~R. Rajamathi, W.~Schnelle, O.~Barkalov,
  M.~Hanfland, S.-C. Wu, C.~Shekhar, Y.~Sun, V.~S{\"u}{\ss}, M.~Schmidt,
  U.~Schwarz, E.~Pippel, P.~Werner, R.~Hillebrand, T.~F{\"o}rster, E.~Kampert,
  S.~Parkin, R.~J. Cava, C.~Felser, B.~Yan, and S.~A. Medvedev,
\newblock Nat. Commun. {\bf 7}, 11038 (2016).

\bibitem{Takahashi2017}
H.~Takahashi, T.~Akiba, K.~Imura, T.~Shiino, K.~Deguchi, N.~K. Sato, H.~Sakai,
  M.~S. Bahramy, and S.~Ishiwata,
\newblock Phys. Rev. B {\bf 95}, 100501 (2017).

\bibitem{Heikes2018}
C.~Heikes, I.-L. Liu, T.~Metz, C.~Eckberg, P.~Neves, Y.~Wu, L.~Hung,
  P.~Piccoli, H.~Cao, J.~Leao, J.~Paglione, T.~Yildirim, N.~P. Butch, and
  W.~Ratcliff,
\newblock Phys. Rev. Mater. {\bf 2}, 074202 (2018).

\bibitem{Hu2019}
Y.~J. Hu, Y.~T. Chan, K.~T. Lai, K.~O. Ho, X.~Guo, H.-P. Sun, K.~Y. Yip,
  D.~H.~L. Ng, H.-Z. Lu, and S.~K. Goh,
\newblock Phys. Rev. Mater. {\bf 3}, 034201 (2019).

\bibitem{Crepaldi2017}
A.~Crepaldi, G.~Aut{\`{e}}s, A.~Sterzi, G.~Manzoni, M.~Zacchigna, F.~Cilento,
  I.~Vobornik, J.~Fujii, P.~Bugnon, A.~Magrez, H.~Berger, F.~Parmigiani, O.~V.
  Yazyev, and M.~Grioni,
\newblock Phys. Rev. B {\bf 95}, 041408(R) (2017).

\bibitem{Tamai2016}
A.~Tamai, Q.~Wu, I.~Cucchi, F.~Bruno, S.~Ricc{\`{o}}, T.~Kim, M.~Hoesch,
  C.~Barreteau, E.~Giannini, C.~Besnard, A.~Soluyanov, and F.~Baumberger,
\newblock Phys. Rev. X {\bf 6}, 031021 (2016).

\bibitem{Weber2018}
A.~P. Weber, P.~R\"u\ss{}mann, N.~Xu, S.~Muff, M.~Fanciulli, A.~Magrez,
  P.~Bugnon, H.~Berger, N.~C. Plumb, M.~Shi, S.~Bl\"ugel, P.~Mavropoulos, and
  J.~H. Dil,
\newblock Phys. Rev. Lett. {\bf 121}, 156401 (2018).

\bibitem{Rhodes2017}
D.~Rhodes, R.~Sch\"onemann, N.~Aryal, Q.~Zhou, Q.~R. Zhang, E.~Kampert, Y.-C.
  Chiu, Y.~Lai, Y.~Shimura, G.~T. McCandless, J.~Y. Chan, D.~W. Paley, J.~Lee,
  A.~D. Finke, J.~P.~C. Ruff, S.~Das, E.~Manousakis, and L.~Balicas,
\newblock Phys. Rev. B {\bf 96}, 165134 (2017).

\bibitem{Chen2018prb}
F.~C. Chen, X.~Luo, J.~Yan, Y.~Sun, H.~Y. Lv, W.~J. Lu, C.~Y. Xi, P.~Tong,
  Z.~G. Sheng, X.~B. Zhu, W.~H. Song, and Y.~P. Sun,
\newblock Phys. Rev. B {\bf 98}, 041114 (2018).

\bibitem{Luo2016}
X.~Luo, F.~C. Chen, J.~L. Zhang, Q.~L. Pei, G.~T. Lin, W.~J. Lu, Y.~Y. Han,
  C.~Y. Xi, W.~H. Song, and Y.~P. Sun,
\newblock Appl. Phys. Lett. {\bf 109}, 102601 (2016).

\bibitem{Zhong2018}
S.~Zhong, A.~Tiwari, G.~Nichols, F.~Chen, X.~Luo, Y.~Sun, and A.~W. Tsen,
\newblock Phys. Rev. B {\bf 97}, 241409 (2018).

\bibitem{Liu2019}
I.-L. Liu, C.~Heikes, T.~Yildirim, C.~Eckberg, T.~Metz, S.~Ran, W.~Ratcliff,
  J.~Paglione, and N.~P. Butch,
\newblock arXiv:1905.02277 .

\bibitem{AChen2016}
A.~Chen and M.~Franz,
\newblock Phys. Rev. B {\bf 93}, 201105 (2016).

\bibitem{AChen2017}
A.~Chen, D.~I. Pikulin, and M.~Franz,
\newblock Phys. Rev. B {\bf 95}, 174505 (2017).

\bibitem{Xu2018}
N.~Xu, Z.~Wang, A.~Magrez, P.~Bugnon, H.~Berger, C.~Matt, V.~Strocov, N.~Plumb,
  M.~Radovic, E.~Pomjakushina, K.~Conder, J.~Dil, J.~Mesot, R.~Yu, H.~Ding, and
  M.~Shi,
\newblock Phys. Rev. Lett. {\bf 121}, 136401 (2018).

\bibitem{SUPP}
See Supplemental Material for (i) method of single crystal growth, (ii) method
  of DFT calculations, (iii) high magnetic field data on $\#$C up to 36 T from
  NHMFL Tallahassee, (iv) angular dependence of SdH oscillations of $\#$A and
  the comparison with calculated frequencies, (v) Lifshitz-Kosevich analysis
  for various quantum oscillation frequencies, (vi) a table listing all
  effective masses extracted, and (vii) a comparison of the MR, SdH 
  spectra, and carrier mobilities of several high-RRR and low-RRR samples. The Supplemental Material includes Refs.~\cite{Schwarz2003,Perdew1996,Rourke2012}.

\bibitem{Schwarz2003}
K.~Schwarz and P.~Blaha,
\newblock Comput. Mater. Sci. {\bf 28}, 259 (2003).

\bibitem{Perdew1996}
J.~P. Perdew, K.~Burke, and M.~Ernzerhof,
\newblock Phys. Rev. Lett. {\bf 77}, 3865 (1996).

\bibitem{Rourke2012}
P.~M.~C. Rourke and S.~R. Julian,
\newblock Comput. Phys. Commun. {\bf 183}, 324 (2012).

\bibitem{Shoenbergbook}
D.~Shoenberg,
\newblock {\em Magnetic oscillations in metals} (Cambridge University Press,
  2009).

\bibitem{Cai2015}
P.~L. Cai, J.~Hu, L.~P. He, J.~Pan, X.~C. Hong, Z.~Zhang, J.~Zhang, J.~Wei,
  Z.~Q. Mao, and S.~Y. Li,
\newblock Phys. Rev. Lett. {\bf 115}, 057202 (2015).

\bibitem{Zhang2017}
J.~L. Zhang, C.~Y. Guo, X.~D. Zhu, L.~Ma, G.~L. Zheng, Y.~Q. Wang, L.~Pi,
  Y.~Chen, H.~Q. Yuan, and M.~L. Tian,
\newblock Phys. Rev. Lett. {\bf 118}, 206601 (2017).

\end{thebibliography}
\providecommand{\noopsort}[1]{}\providecommand{\singleletter}[1]{#1}%

\end{document}